\documentclass[aps,prb,twocolumn,showpacs]{revtex4}
\usepackage{graphicx}

\begin{document}

\title{Interaction of hydrogen molecules with superconducting nanojunctions}

\author{P.~Makk, Sz.~Csonka, A.~Halbritter}
\affiliation{Department of Physics, Budapest University of
Technology and Economics and \\
Condensed Matter Research Group of the Hungarian Academy of Sciences, 1111 Budapest, Budafoki ut 8., Hungary}

\date{\today}

\begin{abstract}
In this paper the interaction of hydrogen molecules with atomic-sized superconducting nanojunctions is studied.
It is demonstrated by conductance histogram measurements that the superconductors niobium, tantalum and aluminum show a strong interaction with hydrogen, whereas for lead a slight interaction is observed, and for tin and indium no significant interaction is detectable. For Nb, Ta and Pb subgap method is applied to determine the transmission eigenvalues of the nanojunctions in hydrogen environment. It is shown, that in Nb and Ta the mechanical behavior of the junction is spectacularly influenced by hydrogen reflected by extremely long conductance traces, but the electronic properties based on the transmission eigenvalues are similar to those of pure junctions. Evidences for the formation of a single-molecule bridge between the electrodes -- like in recently studied platinum hydrogen system -- were not found.
\end{abstract}


\maketitle

\section{Introduction}

The future aim of building functional single-molecule electronic
devices necessitates the development of reliable characterization
techniques of molecular nanojunctions. In many cases the direct
microscopic imaging of the junction is not possible, thus all the
information about the molecular device must be extracted from its
electronic properties. The conduction properties of the device can
be characterized by the so-called mesoscopic PIN-code,\cite{agrait} the set of
the transmission eigenvalues of the junction. According to the
Landauer formula, the linear conductance of the junction is
proportional to the sum of the transmission eigenvalues,
$G=2e^2/h\sum\tau_i$, where $G_0=2e^2/h$ is the quantum
conductance unit. For the determination of the individual
transmission eigenvalues the measurement of further quantities is
required. Conductance fluctuation and shot noise measurements
provide an additional independent combination of the mesoscopic
PIN code, respectively $\sum\tau_i(1-\tau_i^2)$ and
$\sum\tau_i(1-\tau_i)$ can be determined.\cite{Ludoph_condfluct,Brom_shotnoise} These methods were
successfully applied to show that a molecular hydrogen junction
between platinum electrodes has a single, perfectly transmitting
channel.\cite{smit, Djukic_shotnoise} However, for a junction with arbitrary conductance and a
larger number of conductance channels the conductance fluctuation
and shot noise measurements are not efficient.

By placing the junction between superconducting electrodes, and
measuring the nonlinear subgap features in the current voltage
characteristic, principally all the transmission eigenvalues can
be determined. This method was successfully used in the study of
single-atom junctions of Nb, Pb and Al, demonstrating that even
for a junction with $5-6$ conductance channels the transmission
eigenvalues can be determined with high precision.\cite{Scheer,Ludoph_subgap,Landman_Nb}
With the use of
proximity effect the sub-gap method can even be extended for
non-superconducting materials, as shown by the sub-gap
measurements on gold nanojunction.\cite{Scheer_proximity}
According to our knowledge, studies on molecular nanojunctions applying the
superconducting subgap method were not yet reported.

Beside the study of the mesoscopic PIN-code and the corresponding
elastic transmission properties of the junction, a fingerprint of
the inelastic excitations of the molecular device can be given by
point-contact spectroscopy and inelastic electron tunneling
spectroscopy measurements. The excitation of the vibrational modes
of the molecular device,\cite{Djukic_vibrmodes} or the excitation of the molecule to an
other configurations with different binding energy\cite{Halbritter_NDC} are reflected
by step- or even peak-like structures in the differential
conductance, $dI/dV(V)$ characteristics. As the subgap features
appear below the superconducting gap on the millivolt or even
sub-mV scale, whereas the inelastic excitations are typically
observed in the range of $30-100$\,mV, from a single $I-V$ curve
measurement on a molecular junction between superconducting
electrodes not only the mesoscopic PIN-code can be determined, but also the inelastic
excitations can be studied.

In this paper we study the behavior of superconducting atomic-sized
nanojunctions in interaction with the simplest molecule, hydrogen.
As superconducting electrodes Nb, Ta, Al, Pb, Sn, and In samples were
used. For all these materials the interaction is investigated by the conductance
histogram technique, and for the superconductors Nb, Ta and Pb
the subgap curves of the molecular nanojunctions are also studied.
The $I-V$ characteristics of the molecular junctions show clear-cut subgap features,
from which the transmission eigenvalues can be extracted with high
precision.

\section{Experimental method}

Our experiments were performed by the mechanically controllable
break junction (MCBJ) technique at liquid helium temperature
($T=2.2-4.2$\,K). The atomic-sized nanojunctions were created under
cryogenic vacuum, by breaking high-purity wires. The hydrogen molecules were admitted to the sample space
through a solenoid valve of a mass flow controller, from a
container of high-purity hydrogen with reduced pressure ($\sim10\,$mbar). For
the removal of contaminants a liquid nitrogen cold-trap was used.
By applying short pulses of $100$\,ms on the solenoid valve, doses
of $\sim 0.1\,\mu$mol H$_2$ could be admitted the sample holder.
The molecules were directed to the close vicinity of the junction
in the cryogenic part of the setup through a capillary tube. For
the reduction of the noise level the input signal was attenuated
and RC filters were placed close to the junction. According to the
width of the peaks in the $dI/dV(V)$ curves of Nb, Ta and Pb tunnel
junctions the noise level was smaller than $80\,\mu$V.

\section{Conductance histograms and individual conductance traces}

\begin{figure}
\centering
\includegraphics[width=\columnwidth]{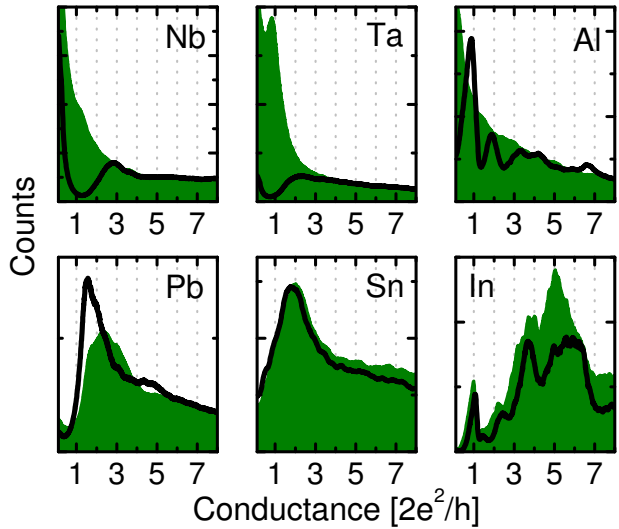}
\caption{\it Conductance histograms of Nb, Ta, Al, Pb, Sn and In junctions
in high vacuum (line graphs) and in hydrogen environment (area graphs).
The histograms were recorded at $T=4.2\,$K and $V=10\,$mV.
All the histograms are normalized to the
precise number of the included traces (4000-8000).}
\label{Histograms.fig}
\end{figure}

Figure \ref{Histograms.fig} shows the conductance
histograms of Nb, Ta, Al, Pb, Sn and In junctions both in high vacuum and
in hydrogen environment. In pure environment the $d$-metals Nb and Ta and the
$p$-metals Pb and Sn exhibit a single broad peak corresponding to single atom junctions, whereas Al shows
more well-defined peaks. All these results agree with previous observations.\cite{agrait} Indium, for which former measurements are not known to us, shows a rather unique histogram: a sharp peak is
observed at the conductance quantum unit and further features are seen at higher conductance values.

In hydrogen environment the histograms of Pb, Sn and In show similar shape as in high vacuum. For lead the peak becomes broader, and shows a significant shift towards higher values; tin does not show any change; whereas for indium the $1$\,G$_0$ peak is reproduced, and the change at higher conductance values is within the scatter of the pure histograms from sample to sample. In contrast for Nb, Ta and Al the addition of hydrogen has a very strong influence on the histograms, in every case a rather featureless distribution is observed demonstrating the appearance of a large number of different hydrogen-related configurations. In Nb frequently a small shoulder near the quantum conductance unit is superimposed on the featureless histogram. In Ta even a small peak at $1$\,G$_0$ can be seen after the admission of hydrogen which disappears by time, as already reported in Ref.~\onlinecite{Shklyarevskii}. It is noted that all the hydrogen-related features disappear if the histograms are recorded at large bias voltage ($V\approx300-400\,$mV), as the hydrogen is desorbed due to the local heating of the junction.

As the histograms are normalized to the number of traces included, the weights in pure and hydrogen environment can be directly compared. Accordingly, at  higher conductance values where the effect of hydrogen is negligible the two histograms fall onto each other. For aluminum the overall weight of the two histograms is similar, just the characteristic peaks of aluminum are smoothed out in the hydrogen affected histogram. In Nb and Ta a remarkable feature is observed: until the conductance of the pure single atom contacts ($G\approx2.3\,$G$_0$) the two histograms are almost the same, but below that value the pure histogram shows a minimum, whereas the in the hydrogen affected case a huge weight is observed.

\begin{figure}[h!]
\centering
\includegraphics[width=\columnwidth]{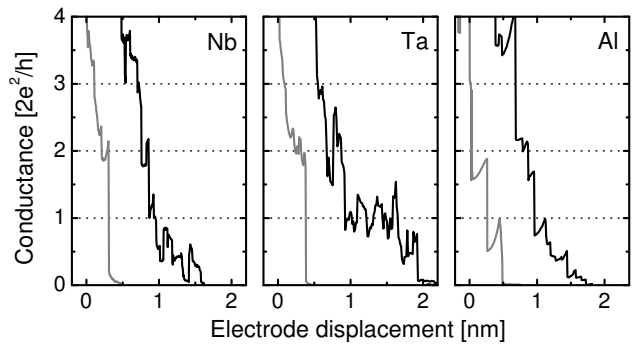}
\caption{\it Typical conductance traces of Nb, Ta, and Al junctions in high vacuum (gray)
and in hydrogen environment (black).}
\label{Condtraces.fig}
\end{figure}

We have found, that the strong interaction of Nb, Ta, and Al with hydrogen is not only statistically detectable by conductance histograms, but it is also clearly visible on individual conductance traces. The gray curves in Fig.~\ref{Condtraces.fig} show
typical conductance traces of pure Nb, Ta and Al, respectively. After forming a
single-atom junction a sudden jump to the tunneling region is observed. In contrast the conductance
traces in hydrogen environment show various different conductance
values with a lot of jumps. The difference is remarkable, just by looking at a single
conductance trace the presence of hydrogen can be detected. In the case of Nb and Ta it is clearly visible that the
hydrogen affected traces are significantly longer than the pure traces. Surprisingly, this kind of traces are not only observed during the rupture of the junction, but the same behavior is seen when the electrodes are pushed together.
In the case of aluminum the characteristic positive slope of the plateaus is preserved in hydrogen environment,
just more smaller jumps are observed instead of the few larger jumps in high vacuum.
For Pb, Sn and In the characteristic shape of the individual conductance traces did not change
by the addition of hydrogen, as expected.

The effect of hydrogen can be further studied by investigating the length of the conductance traces. As
the plateaus length histograms did not show any distinct feature (e.g. chain formation), we characterize the length of the traces with a single number, the average distance between the position of the single atom peak in the pure histogram ($2.3\,$G$_0$ for Nb and Ta and $0.85\,$G$_0$ for Al), and $0.1\,$G$_0$ during the opening of the junction. According to calibration measurements based on the exponential variation of the tunnel current, in pure environment this average length is approximately $1\,$\AA\ for all the three metals. In hydrogen environment the traces of aluminum junctions are slightly longer: the above defined average length is $1.4\times$ larger than that of pure junctions. In niobium and tantalum a huge increase is observed, the hydrogen affected traces are $6-9$ times longer than the pure ones. A similar increase is also observed if the length is measured during the closing of the junctions.

\begin{figure}
\centering
\includegraphics[width=\columnwidth]{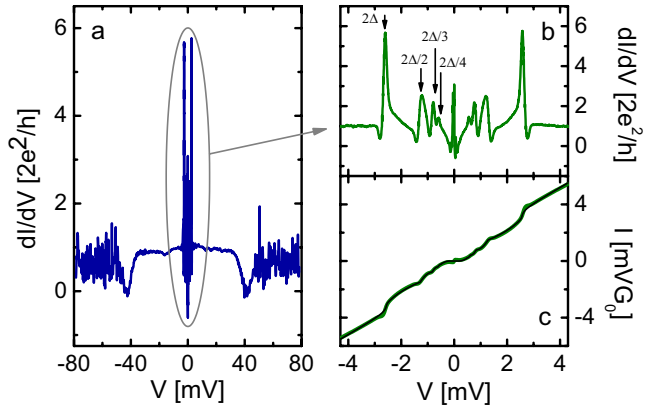}
\caption{\it $I-V$ curve measurement on a hydrogen affected Nb junction with $G\approx1\,G_0$.
(a) The differential conductance in a wide voltage window, showing negative differential conductance phenomenon
and a large conductance noise at higher voltages. (b) Differential conductance in the superconducting gap region. (c) $I-V$ curve in the superconducting gap region and the best theoretical fit corresponding to the set of channel transmissions \{0.70,0.14,0.08,0.05,0.04\}.}
\label{Vibrspectrum.fig}
\end{figure}

\section{Subgap analysis}

\begin{figure}
\centering
\includegraphics[width=0.9\columnwidth]{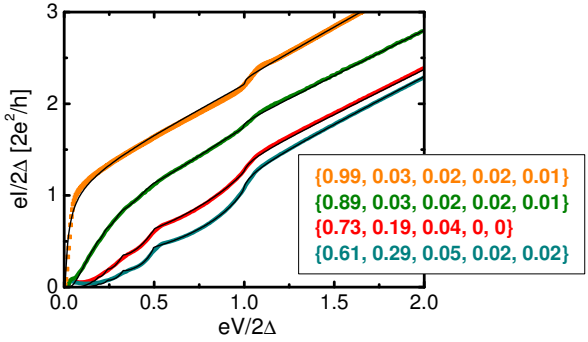}
\caption{\it Subgap structure measurements on 4 different niobium junctions in hydrogen environment with $G\approx1\,G_0$. The set of transmission eigenvalues corresponding to the best fits (thin black lines) are also
indicated on the figure.}
\label{Subgap1G0.fig}
\end{figure}

Fig.~\ref{Vibrspectrum.fig} shows the results of an $I-V$ curve measurement on a
$G=1\,$G$_0$ Nb junction in hydrogen environment. Panel (a) shows the the differential
conductance, $dI/dV$ in a wide voltage window. The curve shows a negative differential
conductance peak at $V=40\,$mV and a large noise at higher voltages. Both features are
characteristic of molecular nanojunctions, and cannot be seen in pure environment.
The negative differential conductance peak is related to the scattering on a two-level
system formed by the molecular junction, but the more precise microscopic background of the
phenomenon is still debated.\cite{Halbritter_NDC,thijssen_VITLS} The observation of a shoulder at the quantum conductance unit in the
histogram of niobium-hydrogen, together with the appearance of a peak-like structure close to
the previously observed vibrational energy in Pt-H$_2$ molecular bridges would imply the formation of
a hydrogen molecular bridge between the niobium electrodes similarly to the Pt-H$_2$ system.\cite{smit,Djukic_vibrmodes}
This hypothesis can be tested by zooming on the superconducting features in the middle of the $I-V$ curve.
Fig.~\ref{Vibrspectrum.fig}b shows the differential conductance within the gap region exhibiting distinct peaks
at the fractional values of the gap due to multiple Andreev reflections.\cite{Scheer,Ludoph_subgap} Fig.~\ref{Vibrspectrum.fig}c shows the $I-V$ curve itself, which is fitted by the theory of multiple Andreev reflections.\cite{Riquelme} The fitting shows that the junction has 5 open channels, and the corresponding transmission probabilities are: $\{0.70,0.14,0.08,0.05,0.04\}$. This result clearly shows that no hydrogen molecular bridge is formed, which would have a single conductance channel.
Figure \ref{Subgap1G0.fig} shows further examples of subgap curves in niobium-hydrogen junctions with $G\approx1\,$G$_0$. It is seen, that occasionally subgap curves with a single, perfectly transmitting conductance channel are indeed observed, but these curves were rather rare, and occasionally similar curves were also observed in pure Nb junctions. In the majority of the cases the second conductance channel also gives a significant contribution.
We note, that in our measurements we did not observe clear vibrational spectra like in the Pt-H$_2$ system,\cite{smit,Djukic_vibrmodes}
in hydrogen environment rather huge peak-like structures and/or large noise of the conductance at higher bias voltage were seen, which are both demonstrated in Fig.~\ref{Vibrspectrum.fig}.

\begin{figure}
\centering
\includegraphics[width=\columnwidth]{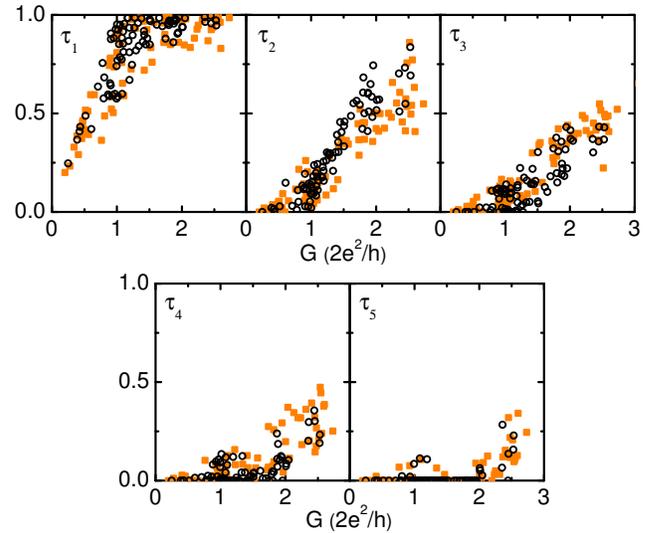}
\caption{\it Transmission probabilities of a large amount of independent Ta nanojunctions in high vacuum (gray squares) and in hydrogen environment (black circles) as a function of conductance.}
\label{Pincode.fig}
\end{figure}

Taking advantage, that with the subgap method the complete mesoscopic PIN-code of the junction can be determined at an arbitrary point of the conductance trace, we have performed a large number of subgap measurements in a wide conductance interval of $0.1\,$G$_0$ - $3\,$G$_0$, both for pure and hydrogen affected Nb, Ta and Pb junctions. Our results show the strange conclusion, that in the whole conductance interval no statistical difference can be pointed out between the channel distribution of the pure and hydrogen affected junctions. This result is demonstrated for Ta junctions in Fig.~\ref{Pincode.fig}, the five panels show the distribution of the transmission probabilities for a large number of independent configurations both in pure and hydrogen environment. It is clear that the opening of the conductance channels follow the same tendency for pure and hydrogen affected junctions, the measured points fall onto each other within the scattering of the data. The same behavior was observed for Nb and Pb junctions as well, for which the results are demonstrated in Table \ref{subgaptable} showing the mean value and the standard deviation of the channel transmissions at some characteristic conductance values. For Nb these values were chosen as $0.3\,$G$_0$ corresponding to a point from a tail of the hydrogen affected histogram with a large weight; $1\,$G$_0$ corresponding to the shoulder in the hydrogen affected histogram; and $2.3\,$G$_0$ corresponding to the peak due to single-atom contacts in the pure histogram. For Pb the results for $1.5\,$G$_0$ and $2.2\,$G$_0$ are presented corresponding to the peak positions in the pure and hydrogen affected histograms, respectively. For comparison the results for Ta junctions with $G=1\,$G$_0$ corresponding to the peak in the hydrogen affected histogram are also shown. It is seen that at each conductance value the mean value of all the transmission coefficients for pure and hydrogen affected junctions are the same well within the standard deviation of the data.

\begin{table}
\centering
\includegraphics[width=\columnwidth]{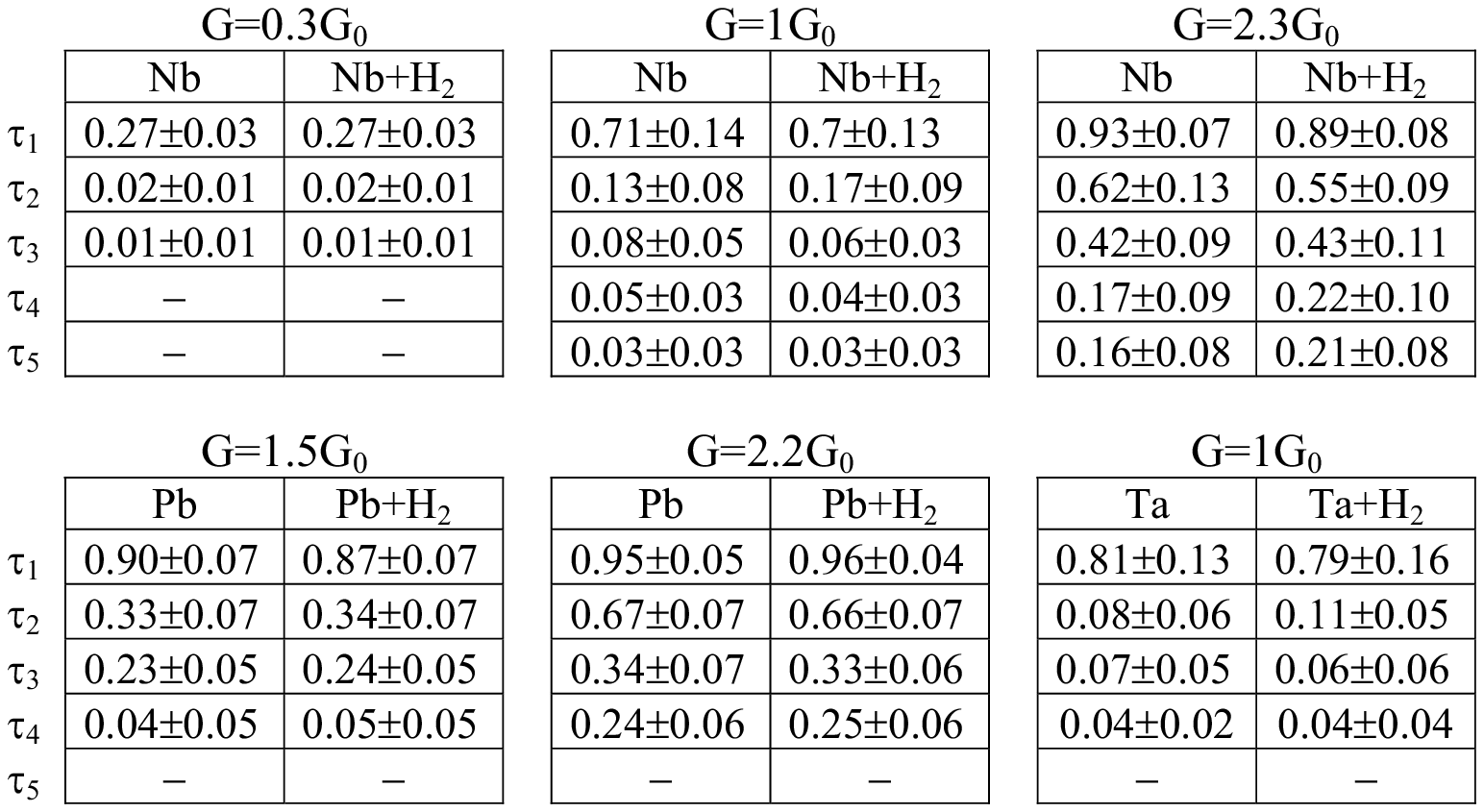}
\caption{\it The mean value and the standard deviation of the transmission eigenvalues at some characteristic conductance values of Nb, Ta and Pb junctions both in high vacuum and in hydrogen environment. The indicated conductance values were adjusted with an accuracy of a $0.03\,G_0$, and at each conductance value the subgap curves were recorded on 30 independent junctions.}
\label{subgaptable}
\end{table}

\section{Summary and discussion of the results}

Our measurements have shown that in Nb, Ta and Al junctions the conductance histograms are strongly reshaped due to the interaction with hydrogen; in Pb some effect of hydrogen is observed, whereas in Sn and In no significant change is observed.

In Al junctions the results imply, that hydrogen stabilizes various arrangements with slightly different conductance values, and so the histogram is smoothed and the individual traces show several small steps. The overall weight of the histogram and length of the traces does not change considerably, and the characteristic shape of the plateaus with positive slope is also preserved. For the subgap analysis of aluminum-hydrogen junctions sub-kelvin temperatures would be required, which is not available in our setup.

In Pb junctions the subgap analysis shows that the transmission probabilities in hydrogen environment are indistinguishable from those of pure junctions. This together with the slight changes in the histogram imply that the interaction of lead with hydrogen is weak.

Niobium and tantalum junctions show a very similar behavior, which is not surprising, as Nb is placed above Ta in the periodic table. Both metals show strong interaction with hydrogen, in H$_2$ environment a rather featureless histogram is observed, in which at low conductance values a huge weight grows compared to the pure histograms. For both metals some features can be seen close to the quantum conductance unit, in Nb occasionally a shoulder is seen in the hydrogen affected histogram, whereas in Ta even a peak can grow at $1\,$G$_0$, which disappears by time. The strong interaction with hydrogen is evident from individual conductance traces, instead of the sudden jump to tunneling extremely long traces with a lot of jumps are observed. The length analysis shows that in hydrogen the average length to break a single atom contact is $6-9$ times larger than in high vacuum. This feature indicates a completely different mechanical behavior of hydrogen affected Nb and Ta junctions. As no periodic peaks are observed in the plateaus' length histogram, and the same extremely long traces are also observed during the closing of the junctions,
the formation of atomic chains is not a plausible explanation for this strange phenomenon. Rather a very soft, plastic behavior of the junction's neighborhood and the pulling of a long neck is expected. The subgap analysis of the hydrogen affected junctions has shown that in spite of the growth of a shoulder or even a peak near the conductance quantum unit in the histograms of Nb and Ta and the frequent occurrence of peak-like structures in the differential conductance curves, the formation of a molecular hydrogen bridge with a single conductance channel is not observed in the large majority of the cases. Furthermore our results show that the mesoscopic PIN-code of the hydrogen affected junctions are the same as those of pure junctions in a wide conductance interval. This indicates that the conduction properties are still dominated by Nb and Ta, though the mechanical behavior of junctions is completely changed in hydrogen environment. All these results imply, that a larger number of molecules interacting with the metallic junction - and probably also being dissolved in the electrodes - cause a soft, plastic behavior of the junctions, but no well-defined single-molecule contacts are formed.

\section{Conclusions}

In conclusion, we have studied the interaction of superconducting metals Nb, Ta, Al, Pb, Sn and In with hydrogen molecules. The analysis of conductance histograms and individual conductance traces shows that Nb, Ta and Al junctions strongly interact with hydrogen. We have also applied the superconducting subgap method to determine the mesoscopic PIN-code of molecular nanojunctions. Our measurements show that in Nb and Ta the interaction with hydrogen completely changes the mechanical behavior of the junctions, but the electronic properties based on the transmission probabilities are the same as those of pure junctions. These results imply, that the interaction with hydrogen causes a soft, plastic behavior of the junctions, but no well-defined single-molecule contacts are formed, and the conduction properties are still dominated by Nb and Ta.

Our measurements also demonstrate, that the break junction method with superconducting electrodes provides a powerful tool for the study of molecular nanojunctions: beside the fine tuning of the electrode separation both the elastic transmission eigenvalues of the junctions can be determined, and the inelastic excitations of the junction can be studied. All these together provide an outstanding amount of information about the studied molecular nanostructures.

\section{Acknowledgements}

This work has been supported by the Hungarian research funds OTKA
F049330, TS049881. A.~Halbritter is a grantee of the Bolyai
J\'anos Scholarship. The authors are grateful to G. Rubio-Bollinger
for the multiple Andreev reflection fitting code.

\end{document}